# TOTAL SOLAR IRRADIANCE VARIABILITY AND THE SOLAR ACTIVITY CYCLE

Probhas Raychaudhuri

*Department of Applied Mathematics, Calcutta University*
*92 A.P.C. Road, Kolkata-700009, India*
*E-mail: probhasprc@rediffmail.com*

It is suggested that the solar variability is due to the perturbed nature of the solar core and this variability is provided by the variability of the solar neutrino flux from the solar neutrino detectors i.e., Homestake, Superkamiokande, SAGE and GALLEX-GNO. The solar neutrino flux in the standard solar model (SSM) was calculated on the assumption of $L_\nu$ (neutrino luminosity) = $L_\gamma$ (optical luminosity) which implies that if there is a change in optical luminosity then solar neutrino flux data will also be changed. An internal dynamo due to the cyclic variation of nuclear energy generation inside the core of the sun is responsible for the solar activity cycle was suggested and thus the internal magnetic field is also variable. Again the changes in the nuclear energy generation induce structural changes that result in variations of the global solar parameters i.e., luminosity, radius and temperatures etc. From the analysis of total solar irradiance (TSI) data during the year from 1970 to 2003 we have found five phases within the solar activity cycle. The first phase (I) starts before two years from the sunspot minimum. The second phase (II) starts at the time of sunspot minimum and phase (III) starts before 2/3 years from sunspot maximum whereas phase (IV) starts at sunspot maximum and fifth phase (V) starts at after 2-3 years from sunspot maximum.

## 1. Introduction

The TSI variation is very important for the understanding of solar internal structure and the solar terrestrial relationships. The TSI is integrated solar energy flux over the entire spectrum which arrive at the top of the atmosphere at the mean sun-earth distance is called solar constant. The radiative output of the sun established the earth's radiation environment and influences its temperature and atmosphere. It has been indicated that small persistent variation in energy flux may play an important role in climate changes. TSI has been monitored from several satellites, e.g. Nimbus 7, Solar maximum mission (SMM), the NASA, ERBS, NOAA9, NOAA 10, Eureca and the UARS (upper atmospheric research satellite) [1] etc. From these observations it reveals that the total solar irradiance varies about a small fraction of 0.1% over solar cycle being higher during maximum solar activity conditions. TSI variation within solar cycle is thought to be due to the changing emission of bright magnetic elements, including faculae and the magnetic network .Thus solar cycle variability may also be related to changes in photospheric temperature, however it is not clear whether this change can be associated with the bright network components[2] .Standard solar model (SSM) are known to yield the stellar structure to a very high degree of precision but SSM cannot explain the solar activity cycle of 11 years, the reason being that these models do not include the temperature and magnetic variability of the solar core.The cyclical variation of the solar cycle is only one of the manifestations of the solar cycle. The basic generation of magnetic field in the sun appears to involve the combined effects of the nonuniform rotation and the cyclonic convection referred to as $\alpha$-$\omega$ dynamo but there is no understanding of rapid diffusion and dissipation of the strong magnetic field that is essential for the operation of $\alpha$-$\omega$ dynamo which is responsible for the solar activity cycle but SSM with $\alpha$-$\omega$ dynamo cannot explain why the sunspot maximum to sunspot minimum takes about 6 – 6.4 years while for sunspot minimum to sunspot maximum it takes about 4 4.6 years. Satellite observations have revealed an 11-year cycle in the total solar irradiance (TSI) with amplitude of about 0.1% that is phase with the solar magnetic activity cycle. The effective temperature of the sun changes, there is evidence that the solar radius changes too. Solar oscillation frequencies are also known to change with time in step with the solar activity cycles. Thus any solar model constructed to try and model the solar activity cycle must explain all these changes- the changes in the global parameters i.e., irradiance, temperature, and radius, as well as changes in the oscillation frequencies. Raychaudhuri[3] suggested that the solar activity cycle is due to the temperature fluctuation in the solar interior and the temperature fluctuation inside the core of the sun can be related with the variation of solar constant. Raychaudhuri [4] suggested an internal dynamo due to the cyclic variation of nuclear energy generation inside the core of the sun ch is responsible for the solar activity cycle.. Raychaudhuri [3, 4] suggested that magnetic field could be originated from the dynamo action associated with the variation of nuclear energy generation inside the core of the sun, which is related to the variation in temperature and composition. Since 1976 Raychaudhuri [5] and others [6] have already demonstrated that the solar neutrino flux varies with the solar activity cycle. In fact Raychaudhuri [7, 8] showed that the solar neutrino flux data is varying with the solar cycle and the solar neutrino flux data exhibit also multiple phases and periods.

In this paper, we will try to show that solar core structural change may be involved to induce the TSI variability and the structural change of the sun is responsible for the phases of the solar activity cycle.

## 2. Standard Solar Model and Solar Neutrino Flux

Raychaudhuri [3, 4, 9] pointed out that Standard solar model (SSM) are known to yield stellar structure to a very good degree of precision but the SSM cannot explain the solar activity cycle, the reason being that SSM does not include temperature and magnetic variability of the solar core. The temperature variability implied a variation of the energy source and from that source of energy magnetic field can be generated which also imply a magnetic variability. In the SSM neutrino flux from the sun was calculated on the assumption of $L_\nu$ (neutrino luminosity) = $L_\gamma$ (optical luminosity), which implies that if there is a change in the optical luminosity then solar neutrino flux will also be changed i.e., neutrino flux will be variable within the solar cycle. In this connection it may be mentioned that Raychaudhuri [3, 4] suggested a perturbed solar model which can account not only the neutrino flux variability but also the solar irradiance variability within the solar cycle. It is shown by Raychaudhuri [3,4] that solar neutrino flux detected by Homestake, Kamiokande-Superkamiokande, SAGE, GALLEX-GNO detectors are variable in nature. The periodicity of the solar neutrino flux is also compatible with the periodicity of the other solar activities i.e., sunspots, solar flares, solar proton events (E>10 MeV) etc. For the support of Raychaudhuri's perturbed solar model we have demonstrated that solar neutrino flux data are fractal in nature [10].

### 3. Total Solar Irradiance Variability

The ultimate source of the solar energy is the nuclear reactions that are taking place inside the core of the sun. The underlying physical mechanisms are not yet well understood. It is shown that empirical models of TSI, solely based on the surface manifestations of solar magnetic activity cannot explain all the aspects of irradiance changes[11]. Identification of the causes of the residual variability in TSI, which is not explained by the effects of sunspots, faculae and the magnetic network, is a difficult problem since global effects may also produce reactions that are taking place inside the core of the sun and the immediate source of energy is the solar surface. From the analysis of the TSI data of the current 23$^{rd}$ solar cycle Toma et al [12] showed a greater increase in TSI for the early phases of this cycle than expected from measurements of the total magnetic flux and traditional solar activity indices, which indicate that solar cycle 23 is weaker than solar cycle 22.

Raychaudhuri [1, 2] suggested a perturbed solar model to account for the irradiance variation with the solar activity cycle which also suggests that solar activity cycle is due to the pulsating character of the nuclear energy generation inside the core of the sun. Endal et al [13] proposed that a variable internal magnetic field should affect all the general parameters of the sun. Lyden and Sofia [14] found that sensible internal magnetic fields variations would perturb the internal structure of the sun and consequently affect all the global parameters. Endal et al.[13] and Lyden and Sofia [14] avoided the temperature fluctuation inside the core of the sun. Solar neutrino flux variation is related to the temperature fluctuation inside the core of the sun in Raychaudhuri's perturbed solar model and the excess energy will be generated at the time of sunspot minimum and this excess energy will be transformed into magnetic energy, gravitational energy and thermal energy etc. below the tacholine. The magnetic energy $3 \times 10^{38}$ ergs will be generated during the solar cycle due to internal dynamo which governs the solar activity cycle and. internal magnetic field is also varying with the solar cycle. Raychaudhuri pointed out [15] that bright network components, active region populations is related to the changes in the solar internal structures. The temperature fluctuation of the solar interior is responsible for the solar activity cycle and is related to the changes in the TSI variations and thus it appears that the changes of the solar surface temperature is responsible for the TSI variations with the solar activity cycle. Kangas and Raychaudhuri [16] suggested that at the end of the expansion phase a close coupling is possible between the solar energy source and the solar surface layers and this is characterized by the dominant activities of the sun. The outer layers of the sun are not held strongly and it is subjected to get an impulse from the solar core. Due to temperature fluctuation inside the solar core, the radiation pressure moving radially outward and their momentum is absorbed by the solar outer layers. Due to angular momentum conservation, the outer layers must be disturbed and the temperature of the outer layers is slightly increased from $T$ to $T + \Delta T$. Thus solar brightness variations are due to the temperature variations in the outer layers of the sun. Taking from the observations the amplitude of the total solar irradiance shifts $\Delta I_r$ for the eleven year period to be roughly i.e., at $I_r = 1365$ W/cm$^2$. We take $\Delta I_r / I_r = 0.001$. If we consider $I_r \propto R^2 T^4$, and take $\Delta R/R = 2 \times 10^{-5}$ (Emilio et al. [17]) then we will get $\Delta T/T = 2.4 \times 10^{-4}$ and this implies an increase in the surface temperature of $1.44^0$ K which is almost equal to the observed values [18].

## 4. Phases of Solar Activity Cycle, Solar Irradiance Variation and Solar Neutrino flux

It was suggested that the solar cycle consists of five phases with different activities [19]. The five phases suggests that the solar activity is multiperiodic in character. The period ranges from 5 months to 11 years. Kangas and Raychaudhuri [16, 19] suggested that the true beginning of the solar activity starts from 2/3 years before epoch of the relative sunspot number minimum. The period can be taken as phase I. The relative sunspot number minimum can be taken as phase II. It is observed that many outstanding events such as large solar flares, large sunspots, higher neutrino flux, high solar proton events (E > 10 MeV), high Forbush decreases of cosmic rays [20] etc. are observed before 2-3 years and after 2-3 years from the relative sunspot number maximum and these phases can be taken as phase III and V respectively. The relative sunspot number maximum time can be taken as phase IV.

### *4.1. Total Solar Irradiance*

In the case of TSI measured by Nimbus –7/ ERB radiometer there are periodicities with high amplitude at 85.3, 51.2 and 42.7 months. The multiperiodicities observed[21] in the TSI variation indicates that there must be multiphases (i.e., five) of TSI during the solar activity cycle. The five phases of TSI is shown in fig.1

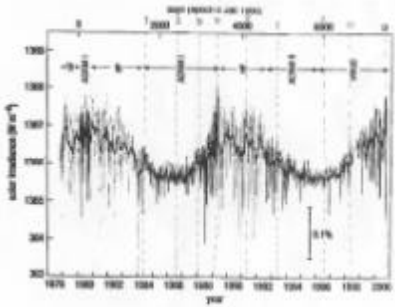

Fig. 1. Composite of total solar irradiance covering more than two solar cycles from 1978 to 2000, as measured by radiometers flying on spacecraft. Data from four instruments have been used to create this composite. Details of the procedure employed to merge the data sets are provided by Frolich and Lean [22]. The five phases are I, II, III, IV, V in the solar cycle.

### *4.2. Solar Neutrino Flux and other Solar Activities*

The observation and interpretation of the TSI may led to new ways of understanding the structure and dynamics of the sun after comparing with the variations of solar neutrino flux, sunspot numbers, low order accoustic p-modes, etc. Homestake solar neutrino flux data shows time variation[23] apart from anticorrelation of sunspot numbers with solar neutrino flux data although there is a controversy of anticorrelation.
between solar neutrino flux data and sunspot data[7,24]. The results of time variation of solar neutrino flux is supported by its correlation with solar diameter and low order acoustic p-modes [24, 25] which may be connected with the variation of the solar inner core temperature with the solar activity cycle. The periodicity around 30 months is observed in solar diameter [26] sunspot data [27], solar neutrino flux data [27]. 32 months periodicity is also present in the full disk magnetic flux. Again short period around 5 to 10 months observed in many solar events such as solar flares, solar proton events (E> 10 MeV), Ca - K plage index, 10.7 cm radio flux and solar radius [14, 19, 20]. The above mentioned properties of the solar events suggest that solar events have multiple periods which indicates that there are multiple phase i.e., five phases in the solar activity cycle. Homestake solar neutrino flux data with five phases is demonstrated in fig.2.

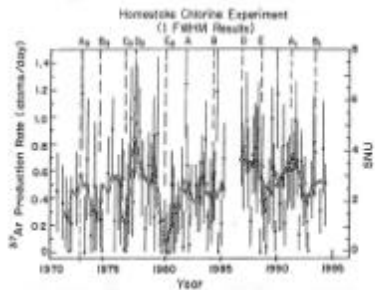

Fig. 2. This plot shows the results for 108 individual solar neutrino observations made with the Homestake Chlorine detector. Five point weighted average data are also shown with the dashed curve. A, B, C, D, E, etc. represent points of five phases of the solar activity cycle.

### *4.3. Five Phases of Solar Neutrino Flux Data from Superkamiokande-I[28] in the Solar Activity Cycle*

Raychaudhuri [9] has shown that there exists five phases in many solar activities i.e., sunspot numbers, solar magnetic fields [29] and solar neutrino flux data in the Homestake, SAGE and GALLEX-GNO detectors. Here we present Superkamiokande–I (SK-I) solar neutrino flux data (six months average from 31 May 1996 to 15 July 2001 (See table-1). We take the solar neutrino flux data 2.289 from July to December 1996 as phase II, 2.468 - 2.489 July to December 1998 to January to June as Phase III and 2.295 – 2.297 from January to June 2000 to July to December 2000 as phase IV and 2.495 at from January 2001 to June 2001 as Phase V and the phase I will occur after 2/3 years from July 2001.Recently Takeuchi [30] have presented SK-1 data from 24 December 2002 to 25 March 2004. The phase I falls from July 2003 to December 2003 and the rate is 2.609. The rate of neutrino flux is lower during sunspot minimum (phase II) and sunspot maximum (phase V) implies strongly that solar neutrino flux is varying with the solar activity cycle. In fact we have found [31] statistically significant periods at 5.2, 11.65, 14.38 and 32.99 months from SK-I solar neutrino flux data. Again SNO group [32] presented time dependence data from 2 November 1999 to 28 August 2003 and have not found statistically significant periodicity in the solar neutrino flux data. The above time period is within phase IV, V and I and we have found similar type of solar neutrino flux counts in the SNO data e.g., higher count in phase IV and I and a lower count in phase V. We suggest that SNO data must have also multiple periods as observed in SK-I [31]. Raychaudhuri [1] suggested that solar total energy spectra is related to the temperature fluctuation of the solar energy source. ACRIM I TSI observations [21] has exhibit periods at 85.3, 51.2 and 42.7 months indicating multiperiodic (multiphase) activities in the solar activity cycle.

Table 1. Superkamiokande-I solar neutrino flux data from 31 July 1996 to 15 July 2001(Six months average data)(DATA/SSM) ( Flux = $10^6$ cm$^{-2}$ sec$^{-1}$ ).

| Year | January to June | July to December |
|------|-----------------|------------------|
| 1996 |                 | 2.289            |
| 1997 | 2.326           | 2.282            |
| 1998 | 2.356           | 2.468            |
| 1999 | 2.489           | 2.425            |
| 2000 | 2.295           | 2.297            |
| 2001 | 2.495           |                  |

## 5. Discussion

The variation of TSI may provide an indication for a secular change that might be related to subtle changes in the solar radius and which may be relating to a pulsating solar core [3, 4, 33]. It is pointed out that [15, 34] evolution of the active region population is also connected with the solar core evolution. Short-term changes of TSI during the solar activity cycles are expected to be due to the luminosity changes connected with the temperature fluctuation of the solar surface and may also be due to the redistribution of the solar radiation by sunspots and active region population. Measurements of TSI varies over a range of periodicities, and most of the observed changes are probably associated with other solar activities e.g., sunspot numbers, solar flares, solar diameter, solar neutrino fluxes, low order acoustic p-modes etc. As the total solar irradiance variations may have impact on the solar – terrestrial relations therefore the phases of the total solar irradiance can have some signature on the terrestrial climate variation. The fluctuations in the sun's radiative output during the solar activity cycle can potentially affect global surface temperatures and influence terrestrial climate and weather, changes the planet's ozone layers etc.